\documentclass{llncs}
 \usepackage{graphicx}
 \usepackage{amsmath}
 \usepackage[width=435pt,height=610pt,centering]{geometry}
 \usepackage{amssymb}
 \usepackage{verbatim}
 \usepackage[algoruled]{algorithm2e}
 \usepackage{xcolor}
 \usepackage{relsize}
 \usepackage[utf8]{inputenc}
 \usepackage{framed}
 \usepackage{dirtree}
\usepackage{array}
 \usepackage[framed]{ntheorem}
\usepackage{dsfont}
\usepackage{enumitem}

 \newcommand{\bigO}{{\mathop{\rm O}}}

\newcommand{\Shuuut}[1]{}

 \newcommand{\Alain}[1]{\Shuuut{\footnote{\color{green}{\bf Alain :} #1{\typeout{#1}}}}}

 \newcommand{\COMMENT}[1]{}

  \newcommand{\Cost}{\text{\sf Cost}}
  \newcommand{\LCost}{\text{\sf LCost}}
  \newcommand{\HeadNum}[1]{{\text{{\sf\bfseries#1}}}}
  \newcommand{\HeadSpa}[1]{\phantom{\HeadNum{#1}}}
  \newcommand{\EqComment}[1]{\colorbox[rgb]{.5,.5,.5}{$\left[\begin{minipage}{.4\textwidth}\relsize{-2}#1\end{minipage}\right]$}\\}
  \renewcommand{\EqComment}[1]{\color[rgb]{.9,.9,.9}\fbox{\color[rgb]{.5,.5,.5}\begin{minipage}{.32\textwidth}\normalcolor\relsize{-2}#1\end{minipage}}\\}

  \newcommand{\A}{\mu}
  \newcommand{\T}{\delta}

  \newcommand{\Def}[1]{{\bfseries{#1}}}
  \newcommand{\ShowProof}[1]{}

\theoremclass{proof}

\theoremstyle{plain}
\theoremsymbol{\Box}
\newshadedtheorem{Proof}{Proof.}

\begin{document}
\mainmatter
\title{Tree decomposition and parameterized algorithms for RNA structure-sequence alignment including tertiary interactions and pseudoknots (extended abstract)}

\author{Philippe Rinaudo\inst{1,2}
      \and
         Yann Ponty\inst{3}
      \and
	 Dominique Barth\inst{2}
      \and
         Alain Denise\inst{1,4}
      }

\institute{LRI, Univ Paris-Sud, CNRS UMR8623 and INRIA AMIB, Orsay, F91405
 	\and PRISM, Univ Versailles Saint-Quentin and CNRS UMR8144 Versailles, F78000
           \and LIX, Ecole Polytechnique, CNRS UMR7161 and INRIA AMIB, Palaiseau, F91128
	\and IGM, Univ Paris-Sud and CNRS UMR8621, Orsay, F91405}

\maketitle

\begin{abstract}
We present a general setting for structure-sequence comparison in a large class of RNA structures that unifies and generalizes a number of recent works on specific families on structures. Our approach is based on {\em tree decomposition} of structures and gives rises to a general parameterized algorithm, where the exponential part of the complexity depends on the family of structures. For each of the previously studied families, our algorithm has the same complexity as the specific algorithm that had been given before. 
 \end{abstract}

\section{Introduction}


The RNA structure-sequence comparison problem arises in two main kinds of applications:
searching for a given structured RNA in a long sequence or a set of sequences,
and three-dimensional modeling by homology.
In~\cite{JiangLMZ02}, Jiang and {\it al.} addressed the problem of pairwise comparison of RNA structures in its full generality.
They defined the edit distance problem on RNA structures represented as graphs, using a set of atomic edition operations.
Notably, they gave a dynamic programming algorithm in $\bigO(nm^3)$ time complexity for comparing a sequence to a  {\em nested} structure
where $n$ is the length of the sequence with known structure and $m$ the length of the sequence of unknown structure.
They also established that computing the edit distance between a sequence and a structure is a Max-SNP-hard problem if the structure contains pseudoknots. 

Meanwhile, consiodering all the known interactions in RNAs including non-canonical ones~\cite{LW2001} and pseudoknots is crucial for precise structure-sequence alignment.
In~\cite{JiangLMZ02}, a polynomial algorithm was developed for pseudoknotted structures, but it involves constraints on the costs of the edition operations.
It has been observed that the so-called {\em H-type} and {\em kissing-hairpin} pseudoknots represent more than $80\%$ of the pseudoknots in known structures~\cite{Rodland2006}.
If the structure contains only only H-Type pseudoknots, the alignment problem can be solved in 
$O(nm^3)$~\cite{Han2008,Wong2011}. Moreover, in~\cite{Han2008} these two classes of pseudoknots where embedded into a more general class, the {\em standard pseudoknots}.
Two $O(nm^k)$ and $O(nm^{k+1})$ algorithms, respectively for a single standard pseudoknot, and for a standard pseudoknot which is embedded inside a nested structure were developed, where $k$ is the so-called {\em degree} of the pseudoknot. Recently, the more general class of simple non-standard pseudoknots was defined, and an algorithm was given in $O(nm^{k+1})$ if alone, or in $O(nm^{k+2})$ for a simple recursion~\cite{Wong2011}.

In the present paper, we give a general setting for sequence-structure comparison in a large class of RNA structures that unifies and generalizes all the above families of structures.
Notably, we handle structures where every nucleotide can be paired to any number of other nucleotides, thus considering all kinds of non-canonical interactions~\cite{LW2001}.
Our approach is based on {\em tree decomposition} of structures and gives rises to a general parameterized algorithm, where the exponential part of the complexity depends on the family of structures. For each of the previously studied families, our algorithm has the same complexity as the specific algorithm that had been given before.
Table~\ref{tableComplex} give a summary of the previous works that are generalized by our approaches, and the time complexity 
of our algorithm for each of the classes.
\begin{table}[t]
{\centering \renewcommand*\DTstyle{\rmfamily}
\begin{tabular}{cccc}
Class of Structures& Time comp. & \begin{minipage}[m]{1.5cm}\centering Multiple \\interactions \end{minipage}& Ref.\\[.7em]
\hline
\begin{minipage}[t]{10cm}
\dirtree{%
.1 RCS -- Recursive Classical Structures\DTcomment{}.
.2 PKF -- Secondary Structures (Pseudoknot-free)\DTcomment{}.
.2 ESP -- Embedded Standard Pseudoknots\DTcomment{}.
.2 SST -- Standard Structures\DTcomment{}.
.3 SPK -- Standard Pseudoknots\DTcomment{}.
.2 2RP -- 2-Level Recursive Simple Non-Standard Pseudoknots\DTcomment{}.
.2 SNS -- Simple Non-Standard Structures\DTcomment{}.
.3 SNP -- Simple Non-Standard Pseudoknots\DTcomment{}.
.2 ETH -- Extended Triple Helices\DTcomment{}.
.3 STH -- Triple Helices\DTcomment{}.
}
\end{minipage}
  & \begin{minipage}[t]{1.8cm}\relsize{+1}\centering
   $O(nm^{k+2})$\\
   $O(nm^3)$ \\
   $O(nm^{k+1})$\\
   $O(nm^k)$ \\
   $O(nm^k)$\\
   $O(nm^{k+2})$\\
   $O(nm^{k+1})$\\
   $O(nm^{k+1})$\\
   $O(nm^3)$\\
  $O(nm^3)$\\
\end{minipage}
  &
  \begin{minipage}[t]{1.5cm}\relsize{+1}
  \centering $\surd$\\
   \quad\\
   \quad\\
   $\surd$\\
   \quad\\
   \quad\\
   $\surd$\\
   \quad\\
   $\surd$\\
   $\surd$\\
  \end{minipage}
   &
 {\begin{minipage}[t]{.7cm}\relsize{+1}\centering --\\
  \cite{JiangLMZ02}\\
   \cite{Han2008}\\
   \cite{Han2008}\\
    --\\
   --\\
   \cite{Wong2011}\\
     --\\
  \cite{Wong2011}\\
   \cite{Wong2012}\\
\end{minipage}}
\\
\hline
\end{tabular} \\}
\label{tableComplex}
\caption{Summary of existing algorithms for structure-sequence alignment.
	  Our approach unifies all these algorithms and, for each class of structures captured by pre-existing works, specializes into time complexities that matches previous efforts:
	  the tree structure represents an inclusion relation. Hence the root class RCS includes all other classes.
	  Notation: $k$ is the degree of the pseudoknot/(simple) standard structure.
          }
\vspace{-0.6cm}
\end{table}

\section{Sequence-Structure Alignment}

At first let us state some definitions, starting with the concept of arc-annotated sequence, which will be used as an abstract representation for RNA structure.

\begin{definition}[Arc-annotated sequence]
 An arc-annotated sequence is a pair $(S,P)$, where $S$ is a sequence
 over an alphabet $\varSigma$ and $P$ is a set of unordered pairs of
 positions in $S$.
\end{definition}

\begin{figure}[t]
\centering
\begin{minipage}{0.40\textwidth}
\centering
 \includegraphics[width=\textwidth]{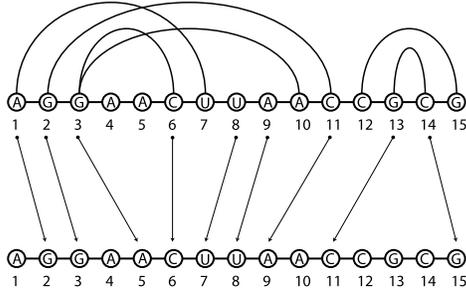}
 \label{alignment}
\end{minipage}
\begin{minipage}{0.55\textwidth}
 \caption{A representation of a partial mapping between an arc-annotated sequence (upper part) and a (plain) sequence (lower part).}
\end{minipage}
\vspace{-1cm}
\end{figure}

For RNA structures, obviously $S$ is the nucleotide sequence and $P$ the set of the interactions over $S$, as illustrated by Figure~\ref{alignment} (Upper part).
Here $\varSigma=\{A,U,G,C\}$, and any $(i,j) \in P$ represents an interaction between the nucleotides at position $i$ and $j$.
The nucleotides are numbered from $1$ to $n$ (where $n$ is the sequence length), follow a 5' to 3' order, and $S[i]$ denotes the nucleotide number $i$.
One should notice that, unlike most definitions, a position $i$ can be involved in multiple interactions, allowing for the representation of tertiary structures.
In the following, we sometimes refer to such an arc-annotated sequence  as an \Def{RNA graph interaction structure} or, in short, an \Def{RNA graph}.

There exists several equivalent ways to define a structure-sequence alignment, that is an alignment between an arc-annotated sequence and a (plain) sequence.
We choose to abstract an alignment as a mapping, i.e. we represent an alignment as a partial mapping between positions in the arc-annotated sequence and positions in the plain sequence, as shown in Figure~\ref{alignment}.

\begin{definition}[Structure-Sequence Alignment]
\label{NewDefAlign}
  A structure-sequence alignment between an arc-annotated sequence $A=(S_A,P_A)$ of size $n$ and a plain sequence $B=(S_B,\varnothing)$ of size $m$ is a partial mapping $\mu$ from $[1,n]$ to $[1,m+1]$ such that:\setlist{nolistsep}
\begin{itemize}
 \item $\mu$ is injective.
 \item $\mu$ preserves the order: $\mu(i) < \mu(j) \Rightarrow i<j$.
\end{itemize}
\end{definition}
We write $\mathcal{F}(A,B)$ (or $\mathcal{F}$ when there is no ambiguity) for the set of all possible alignments between $A$ and $B$.

Remark that some positions in the arc-annotated sequence may be without a corresponding position in the plain sequence.
We note $\mu(i) = \perp$ if position $i$ does not have an image by $\mu$, and qualify it as \Def{unmatched}.
Consecutive sequences of unmatched positions in the structure ($\mu(i) = \perp$) or in the sequence ($\mu^{-1}(i) = \perp$) are usually grouped and scored together.
A \Def{(composite) gap} is then the maximum set of consecutive positions ($i,\dots{},j$) that are either unmatched or do no have an antecedent by $\mu$. The \Def{length} $|g|$ of a gap $g$ is simply the number of positions it contains.
By grouping unmatched positions within gaps, one can handle affine penalty functions in the cost function.

Let us now define the cost of an alignment, which captures the level of similarity between two RNAs, and needs account both for structure and sequence elements.

\begin{definition}[Cost of a Sequence-Structure Alignment] \label{def:cost}
 The cost of a structure-sequence alignment $\mu$ between an arc-annotated sequence $A=(S_A,P_A)$ of size $n$ and a plain sequence $B=(S_B,\varnothing)$ is defined by:
  \begin{equation}\Cost(\mu) = \sum_{\substack{i \in [1,n], \mu(i) \neq \perp}} \gamma(i,\mu(i)) + \sum_{\substack{\text{gap }g\subset A}} \lambda_A(|g|) + \sum_{\substack{\text{gap }g\subset B}} \lambda_B(|g|) + \sum_{\substack{ (i,j) \in P_A }} \varphi(i,j,\mu(i),\mu(j))
  \label{eq:cost}\end{equation}
  where\setlist{nolistsep}
\begin{itemize}
\item $\gamma(i,\mu(i))$ is the cost of a \Def{base substitution} between position $i$ in $A$ and position $\mu(i)$ in $B$.
\item $\lambda_Y(x)=\alpha_Y\cdot x+\beta_Y,$ is the \Def{affine cost penalty} for a gap of length $x$ in a sequence $Y$.
\item $\varphi(i,j,\mu(i),\mu(j))$ is the cost of an \Def{arc removing} ($\mu(i) = \mu(j) = \perp$), \Def{arc altering} ($\mu(i) \text{ or } \mu(j) = \perp $) or \Def{arc substitution}
involving paired positions $i$ and $j$ in $A$.
\end{itemize}
\end{definition}

Note that unlike the score functions defined in \cite{Han2008,Wong2011}, our formulation captures arc-alterations and arc-breakings as atomic operations on their interactions (as in \cite{JiangLMZ02}), allowing for general cost schemes.

\begin{definition}[Structure-Sequence Alignment Problem]
 Given an arc-annotated sequences $A$ and a plain sequence $B=(S_B,\varnothing)$, the structure-sequence alignment problem is to find an alignment between $A$ and $B$ having the minimum cost.
\end{definition}

As stated in \cite{Zhang1999,Blin2010} the problem is already NP-Hard if we consider {\sc Crossing} interactions (\textit{i.e.} pseudoknots, without multiple pairings).
In the following, we will use a parameterized approach to handle general structures including unrestricted crossing interactions and multiple interactions per position, assuming that the total number of interactions per position is bounded by a constant.


\section{Tree Decomposition and Alignment Algorithm}

As previously sketched, our alignment algorithm relies on tree decomposition of an arc-annotated sequence.
Tree decompositions are usually defined on graphs rather than on arc-annotated sequences.
Here we give a straightforward adaptation that preserves all the properties of the standard tree decompositions.

\subsection{Definitions}
\begin{definition}[Tree Decomposition of an arc-annotated sequence]
 Given an arc-annotated sequence $A=(S,P)$, a tree decomposition of $A$ is a pair $(X,T)$ where $X=\{X_{1}, \dots ,X_{N} \}$
 is a family of subsets of positions $\{i,i\in [ 1,n ] \}$, $n=length(S)$,
 and $T$ is a tree whose nodes are the subsets $X_{l}$ (called {\em bags}), satisfying the following properties:\setlist{nolistsep}
\begin{enumerate}
 \item Each position belongs to a bag: $\bigcup_{l\in [ 1,N ]} X_{l} = [1,n]$.
 \item Both ends of an interaction are present in a bag: $ \forall (i,j) \in P$, $\exists l\in [ 1,N ], \{i,j\} \subset X_{l}$.
 \item Consecutive positions are both present in a bag: $ \forall i \in [1,n-1] $, $\exists l \in [ 1,N ], \{i,i+1\} \subset X_{l}$.
 \item For every $X_{l} $ and $X_{s}$, $l, s \in [ 1,N ]$, $X_{l} \cap X_{s}$ $\subset X_{r} $ for all $X_{r}$ on the path between $X_{l}$ and $X_{s}$.
\end{enumerate}
\label{def:treeDec}
\end{definition}

\begin{figure}[t]
\centering
\begin{minipage}{0.57\textwidth}
\centering
  \includegraphics[width=\textwidth]{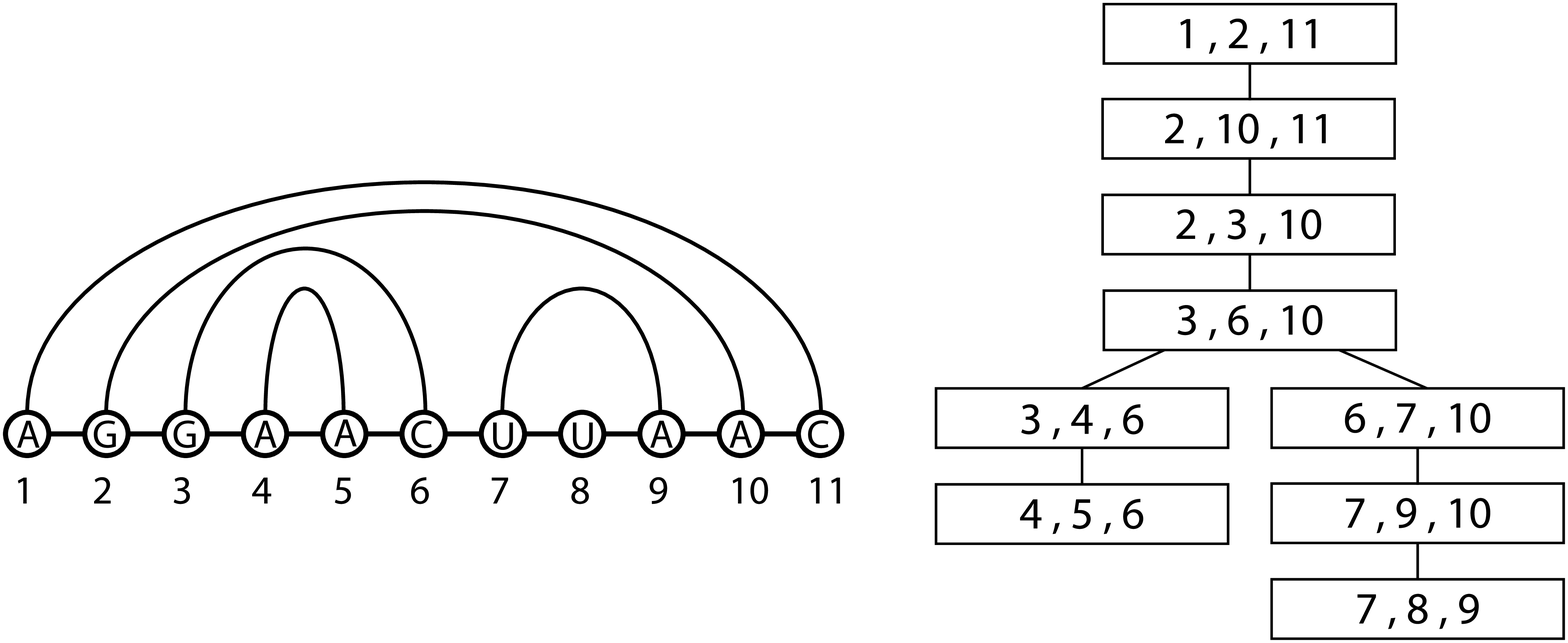}
\end{minipage}
\begin{minipage}{0.40\textwidth}
 \caption{An arc-annotated sequence and an associated tree decomposition.
	  Each bag contains three positions, so the tree width of this tree decomposition is 2.
	  As there is only one bag which does not respect the smooth property (the one with two children), the tree decomposition is 1-weakly-smooth.}
 \label{tree_dec_def}
\end{minipage}
\vspace{-1cm}
\end{figure}

Figure~\ref{tree_dec_def} illustrates the tree decomposition of an arc-annotated sequence.

\begin{definition}[Treewidth]
 The width of a tree decomposition $(X,T)$ is the size of its largest set $X_{l}$ minus one. The \Def{treewidth} $t_w(A)$ of an arc-annotated sequence $A$ is the minimum width among all possible tree decompositions of $A$.
\end{definition}

In general, tree decomposition are not rooted. Nevertheless, for the sake of clarity, we will arbitrarily choose a root $X_0$ in our decompositions.
Additionally, we use the following notation: for any two bags $X_l$ and $X_r$ in $X$, we write $X_{l,r}$ as shorthand for $X_l \cap X_r$.

The process of assigning consecutive positions can be simplified by the affine gap penalties, using the general idea of Gotoh's algorithm~\cite{Gotoh1982}.
Let us then define the notion of smooth bag, that will help us take advantage of this optimization.

\begin{definition}[Smooth Bag of a Tree Decomposition]
\label{def_smooth}
  Let $X_l\in X$ be a bag in a tree decomposition $(X,T)$ for an arc-annotated sequence $A=(S,P)$.
  If $X_l\neq X_0$, then let $X_r$ be its father. $X_l$ is then \Def{smooth} iff
  there exist two consecutive positions $i$ and $j$ such that $i\in X_l-X_r$, $j\in X_{l,r}$, $j$ is not in one of the child of $X_l$, and there is no $i'\in X_l-X_r$ such that $(i',j) \in P$ or $i'$ (except $i$) consecutive to $j$.
  The root $X_0$ is \Def{smooth} iff $(\textit{1})$ there exist two consecutive positions $i,j \in X_0$ such that $j$ is not in any child of $X_0$ and  $(i,j) \notin P$, or $(\textit{2})$ iff the size of the root is strictly smaller than the size of one of its children.
\end{definition}

\begin{definition}[(Weakly-)Smooth Tree Decomposition]
  A tree decomposition $(X,T)$ for an arc-annotated sequence $A=(S,P)$ is \Def{smooth} iff every bag $X_l\in X$ are smooth.\\
  A tree decomposition $(X,T)$ is \Def{$k$-Weakly-Smooth} iff at most $k$ of its bags are not smooth.
\end{definition}

As stated in \cite{Bodlaender1997}, a tree decomposition can always be converted into a binary tree in linear time.
Moreover, this transformation can be done without breaking the smoothness of the tree decomposition.
Therefore, we will limit, without loss of generality, the scope of our algorithm and analysis to binary tree decompositions.
At last, if the tree decomposition is smooth and root $X_0$ is smooth by condition $(1)$, then the tree can made smooth by condition $(2)$ by creating a new root composed of the positions of the old one except the position $i$ (with $i$ as in the definition above) and add it as the father of $X_0$. In the following we will only consider case $(2)$ for the alignment algorithms.

\subsection{An Alignment Algorithm based on Tree Decomposition}
Let us describe an algorithm that computes the minimum cost alignment of an arc-annotated sequence $A$ and a (flat) sequence $B$ ($P_B = \varnothing$).
This algorithm implements a dynamic programming strategy, based on a $k$-Weakly-Smooth tree decomposition of $A$. 

\noindent{\bfseries An alternative internal representation for alignments.}
The recursive step in our scheme consists in extending a partial alignment, assigning positions that are \Def{proper} to the bag (i.e. not present in the bag's father). One of the main challenge is to preserve the sequential order. Indeed, only consecutive  positions $(i,i+1)$ are guaranteed to be simultaneously present in a bag, allowing for a direct control over the sequential ordering ($\A(i)<\A(i+1)$) at the time of an assignment. Intuitively, this property may extend transitively over $\mu$, since $\A(i)<\A(i+1)$ and $\A(i+1)<\A(i+2)$ implies $\A(i)<\A(i+2)$. However, this property  no longer holds when a position is unmatched, as $\A(i+1)$ is then undefined and cannot serve as a reference point for the relative positioning of $\A(i)$ and $\A(i+2)$.

\begin{figure}[t]
    \begin{minipage}[c]{0.35\linewidth}
     {\includegraphics[width=\textwidth]{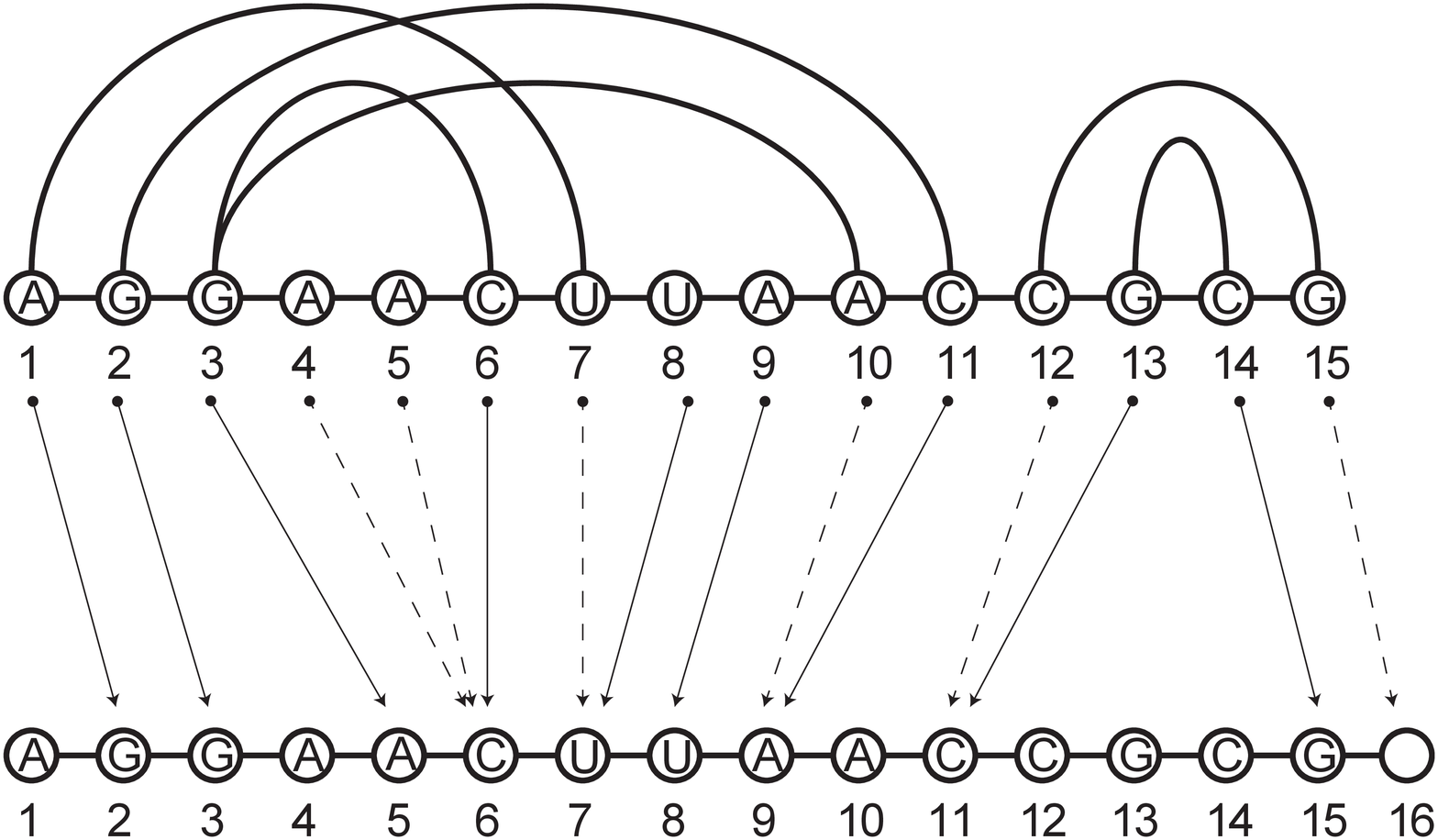}}
    \end{minipage}\hfill
    \begin{minipage}[c]{0.63\linewidth}
      \caption{Internal representation of alignments as a pair $f=(\A,\T)$, used within our dynamic programming algorithm. 
A function $\A$ defines a complete mapping, while $\T$ discriminates matched positions (solid stroke) from unmatched ones (dashed stroke). 
Additionally, unmatched positions are forced to aggregate to their nearest matched neighbor to their right, and we create an additional virtual position (16 here) to
provide an image to trailing unmatched positions in the structure.}\label{fig:alignBis}
    \end{minipage}
\vspace{-1cm}
\end{figure}

To work around this issue,  we identify, in the description of our algorithm, an alignment with a pair $f=(\A,\T)$, where $\A:[1,n]\to[1,m+1]$ is a full ordered mapping ($\A(\cdot)\neq\perp$) and  $\T:[1,n]\to[0,1]$ distinguishes between matched ($\T(i)=1$) and unmatched positions ($\T(i)=0$). As illustrated by Figure~\ref{fig:alignBis}, this new representation aggregates consecutive unmatched positions in $A$ to their nearest rightward position that is matched in the alignment. A \emph{virtual position} $m+1$ is then added to $B$ to serve as an image for the unmatched positions appearing at the end of $A$.  

\noindent{\bfseries Dynamic programming recursion.} Our dynamic programming algorithm assigns positions  in $B$ to the elements of a bag, proceeding to a recursive call that assigns the positions found further down in the tree decomposition.  This requires a few additional notions and definitions.

Let $(S,P)$ be an arc-annotated sequence, and $S'$ be a subset of $S$. The \Def{arc-annotated subsequence induced by $S'$} is the pair $(S',P')$ such that $P'$ is the subset of arcs in $P$ whose both extremities are in $S'$.
  Let $X_l$ be a bag in the tree-decomposition of $A$.
   The \Def{descending subsequence} of $X_l$ is the subset of positions appearing in the descendants of $X_l$ in the tree.
   The \Def{descending arc-annotated subsequence} of $X_l$ is the arc-annotated subsequence induced by its descending subsequence.
The notion of alignment between $A$ and $B$ naturally extends to alignments involving a sub-arc-annotated structure of $A$, and we denote by $\mathcal{F}|_{S'}$ the set of all possible alignments between the arc-annotated subsequence induced by $S'$, and $B$.

Let $X_l$ be a bag having father $X_r$ ($X_r:=X_l$ when $l=0$), let $f \in \mathcal{F}|_{X_{l,r}}$ be an alignment for the common positions of $X_l$ and $X_r$ to $B$. Let us denote by $C^l_f$ the cost of the best alignment between the descending arc-annotated subsequence of $X_l$ and $B$, which matches $f$ on $X_{l,r}$. It can be shown that  $C^l_f$ obeys
\begin{equation}
  C^l_f = \min_{\substack{f'=(\A',\T') \in \mathcal{F}|_{X_l}\\ f'(i)=f(i), \forall i\in X_{l,r} }} \left\{ \LCost(X_l,f') + \displaystyle\sum_{s\in sons(l)}  C^s_{f'|_{X_{s,l}}} \right\}.
\end{equation}
Moreover, the local contribution $\LCost(X_l,f)$ of a bag $X_l$ to an alignment $f=(\A,\T)$ is 
\begin{align*}\LCost(X_l,f) &= \sum_{\substack{i\in X_l-X_r\\ \T(i)=1}} \gamma(i,\A(i)) 
+ \sum_{\substack{  i,j\in X_l \text{ s.t. }\\i\text{ or }j \in X_l-X_r\\\text{ and } (i,j)\in P_A }} \varphi(i,j,f(i),f(j)) \tag*{\relsize{-1} // Bases and interactions} \\
&+ \sum_{\substack{i,i+1\in X_l \text{ s.t. } \\ i\text{ or }i+1\in X_l-X_r \\ \text{and }\A(i+1)>\A(i)}} \alpha_B\cdot(\A(i+1)-\A(i))+\beta_B, \tag*{\relsize{-1} // Gaps in sequence $B$} \\
& + \sum_{\substack{i,i+1\in X_l \text{ s.t. } \\ i\text{ or }i+1\in X_l-X_r \\ \T(i)=1\text{ and }\T(i+1)=0}}\beta_A + \sum_{\substack{i\in X_l-X_r\\\text{s.t. }\T(i)=0}}\alpha_A, 
\tag*{\relsize{-1} // Gaps in sequence $A$} 
\end{align*}
assuming a gentle abuse of notation, in which $f(i)=\{\A(i)\text{ if }\T(i)=1, \text{ or }\perp\text{ otherwise}\}$.

A dynamic programming algorithm follows from this general recurrence equation, in a standard way.
The cost of the best alignment is given by $min\{C^0_f\;|\;f\in \mathcal{F}|_{X_0}\}$.
A simple backtrack procedure gives the best alignment between $A$ and $B$.
The following theorem gives the worst-case complexity of the algorithm.

\begin{theorem}
  Let $A$ and $B$ be two arc-annotated sequences $(P_B =  \varnothing)$,
  and let $(X,T)$ be a tree decomposition of $A$.
  The structure-sequence alignment of $A$ and $B$ can be computed
  in $\Theta(N\cdot m^{t+1})$ time and $\Theta(N\cdot m^t)$ in space,
  where $N=|X|$, $t$ is the tree-width of $(X,T)$, and $m=|B|$.
\label{alignTheo}
\end{theorem}

\ShowProof{\begin{Proof}
  We count each addition as an elementary operation.
  For a given bag $l$, we have $t'=Card(X_{l,r})$ positions to align with the alignment functions in $\mathcal{F}|_{X_{l,r}}$.
  Each position can be aligned with at most $2m+1$ possibilities.
  So there are at most $2m \choose t'$ possible alignments $f \in \mathcal{F}|_{X_{l,r}}$.
  For each of them, a minimum is searched for among all possible alignments of $t''=t+1-t'$ positions.
  This leads to $2m \choose t''$ possible alignments at most.
  Each minimum is computed with at worst $O(t'')$ additions as we stated that there are a constant maximum number of interaction per position and $O(1)$  additions of sons cost.
Thus the time complexity is in $\bigO(m^{t+1})$.
Since we have $N$ bags the time complexity of the alignment is $O(N\cdot m^{t+1})$. \\
Regarding the space complexity,
one needs to create, for each bag, a table of size $\mathcal{O}\left( {2m \choose t'} \right)$ to memorize the best cost for any prior assignment to $X_{l,r}$.
In the worst case, $t'=t$ for all bags (if $t'=t+1$ for some bag, one can discard it since it is totally included in its father).
 Since there are $N$ bags, the space complexity is $\mathcal{O}(N\cdot m^t)$.\qed
\end{Proof}}

This complexity can be further improved when the tree decomposition is smooth (or even only $k$-weakly smooth), by taking advantage of the affine nature of gap penalty functions.
To that purpose, one uses the general principle underlying Gotoh's algorithm~\cite{Gotoh1982}, and introduce a secondary matrix to distinguish gap openings from gap-extensions.
We obtain the dynamic programming equation summarized in Figure~\ref{fig:dptchik}
\begin{figure}[t]
\begin{align*}
\HeadNum{\relsize{+2}1.} && C^l_f &=
   \min\left\{ \begin{array}{c}
    \EqComment{\HeadNum{A} If $\T(i-1)=0$: Aggregate $i$  on $i-1$ ($\A'(i)=\A(i-1)$).\\
    \HeadNum{B}  If $\T(i-1)=1$: $i$ is either next to $i-1$ ($\A'(i)=\A(i-1)+1$, no gap)\ldots
    }
    \displaystyle\min_{\substack{f'=(\A',\T') \in \mathcal{F}|_{X_l}\\\text{s.t.} f'=f \text{ on } X_{l,r}}}
    \Delta_l(f')
    \\
    \EqComment{
    \HeadNum{C} \ldots or further away:  Open a gap and shift $i-1$ to avoid \emph{testing} every position for $i$.
    ($\A''(i-1)=\A(i-1)+1$, and $f'' = f$ otherwise)}
    \alpha_B + \beta_B + D^l_{f''}
  \end{array} \right.
  &
D^l_f &=
   \min\left\{ \begin{array}{c}
    \EqComment{\HeadNum{D} Found a position for $i$ ($\A'(i)=\A(i-1)+1$).}
    \displaystyle\min_{\substack{f'=(\A',\T') \in \mathcal{F}|_{X_l}\\\text{s.t.} f'=f \text{ on } X_{l,r}}}
    \Delta_l(f')
    \\
    \EqComment{
    \HeadNum{E}  Look one step further for a suitable position for $i$ ($\T(i-1)=1$, $\A''(i-1)=\A(i-1)+1$)}
    \alpha_B + D^l_{f''}
  \end{array} \right.\\
\HeadNum{\relsize{+2}2.}&&C^l_f &=
   \min\left\{ \begin{array}{cl}
    \EqComment{
    $\HeadNum{A'}$ If $\T'(i)=0$: Aggregate $i$ on $i+1$ ($\A'(i)=\A(i+1)$).\\
    $\HeadNum{B'}$  If $\T'(i)=1$: $i$ can be right before $i+1$ (no gap)\ldots
    }
    \displaystyle\min_{\substack{f'=(\A',\T') \in \mathcal{F}|_{X_l}\\\text{s.t.} f'=f \text{ on } X_{l,r}}}
    \Delta_l(f')
    \\
    \EqComment{
    \HeadNum{C'} \ldots or further away:  Open a gap and \emph{virtually} shift $i+1$ to avoid \emph{testing} every candidate position for $i$. ($\A''(i+1)=\A(i+1)-1$, and $f'' = f$ otherwise)}
     \alpha_B + \beta_B  + D^l_{f''}
  \end{array} \right.
  &
  D^l_f &= min \left\{
  \begin{array}{cl}
    \EqComment{
    $\HeadNum{D'}$  If $\T'(i)=1$: Found position for $i$ ($\A'(i)=\A(i+1)-1$) $\Rightarrow$ No added gap}
    \displaystyle\min_{\substack{f'=(\A',\T') \in \mathcal{F}|_{X_l}\\\text{s.t.} f'=f \text{ on } X_{l,r}}}
    \Delta_l(f')
    \\
    \EqComment{$\HeadNum{E'}$ If $\T(i+1)=1$, $\A''(i+1)=\A(i+1)-1$,
    and $f'' = f$ otherwise: \emph{Virtually} shifting $i+1$
    $\HeadSpa{E'}$ to avoid \emph{testing} every candidate position for $i$.}
    \alpha_B + D^{l}_{f''}
\end{array} \right.
\\
   &&& \Delta_l(f):=\LCost(X_l,f) + \displaystyle\sum_{s\in sons(l)}  C^s_{f|_{X_{s,l}}}
\end{align*}
\caption{Dynamic programming equation for aligning a smooth bag $X_l$, whose father $X_r$ has previously been assigned. 
Case \HeadNum{1.} and \HeadNum{2.} above apply respectively to smooth bags such that $i-1\in X_r$ or $i+1\in X_r$ (note that these two cases are mutually exclusive from the definition of smoothness). }\label{fig:dptchik}
\end{figure}

\begin{theorem}
  If the tree decomposition of $A$ is $k$-weakly smooth, then the sequence-structure alignment of $A$ and $B$, can be computed in $\Theta(k\cdot m^{t+1} + (N-k)\cdot m^t)$ time.
\label{alignTheoSmooth}
\end{theorem}

\ShowProof{\begin{Proof}
 Computing any of the scores $C^l_f$ or $D^l_f$ requires smoothing bag may only be aligned to a constant number of positions, there only exists $O(m^t)$ candidate alignments for a smooth bag.
 Because there are $N-k$ smooth bags, the final complexity is $\Theta(k(m)^{t+1} + (n-k)m^t)$.
\end{Proof}}

\begin{corollary}
  Let $A$ and $B$ be as before. If the tree decomposition $(X,T)$ of $A$ is smooth and has width $t$, then time complexity of  the structure-sequence alignment algorithm is in $\Theta(N\cdot m^t)$.
\label{alignCoroSmooth}
\end{corollary}

\section{Tree Decomposition and Sequence Structure Alignment of RNA Structures}

\begin{figure}[t]
      \centering
     \includegraphics[width=\textwidth]{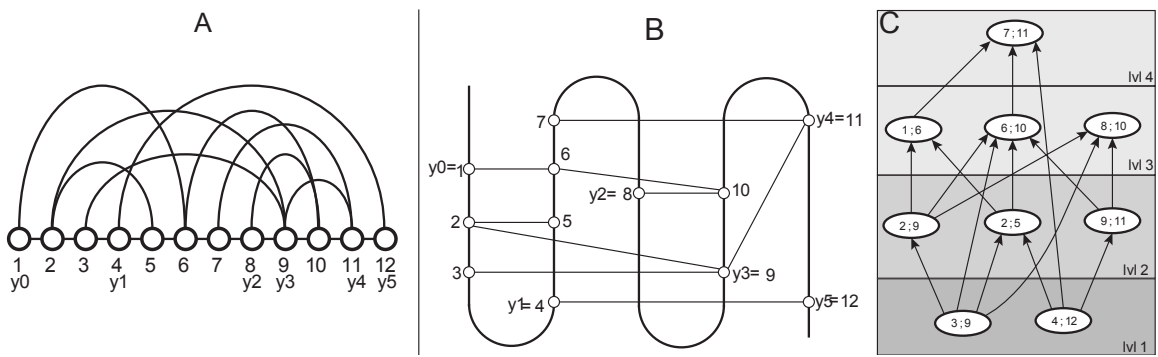}
      \caption{A: An arc-annotated sequence with its pivots. B: An embedding wave representation of the same arc-annotated sequence. C: The associated (acyclic) upward graph (interactions are rank by their level).}
      \label{upward}
\end{figure}

In its full generality, the problem of computing a tree decomposition of minimum width for an arc-annotated sequence is NP-Hard~\cite{Arnborg1987}. However, by restricting the problem to some specific RNA structure families, one can obtain a tree decomposition with a small width in reasonable time. The key idea relies on a total ordering of the positions in the arc-annotated sequence, as shown in the following.
For the sake of clarity, we suppose at first that there is no unpaired position in the structure.
\begin{definition}
 A \Def{wave embedding} $W$ of an arc-annotated sequence $A=(S,P)$ is defined by 
an increasing sequence of \Def{pivot} positions $\mathbf{y}=\{y_i\}_{i=0}^{k}$, such that $y_0:=1$ and $y_k:=n$ (Figure~\ref{upward}). The \Def{degree} of a Wave Embedding is its number of pivots minus one.
\end{definition}
Now, a total ordering on the interactions can be inferred from a wave embedding, in which case the wave embedding is said to be \Def{ordering}. 
Let us now give a sufficient condition for a given embedding to be ordering. To that purpose, we  introduce the \Def{upward graph}, and show that its acyclicity is a sufficient condition for the embedding to be ordering. 



Given a wave embedding $W$ of an arc-annotated sequence $A=(S,P)$, we call \textbf{intervals of $W$} the half open intervals: $I_t=[y_t,y_{t+1}[$ for $t \in [ 1,k-2 ]$ and the interval $I_k=[y_{k-1},y_k]$.
Now, let us  start by defining a partial order $\cdot\prec\cdot$ on the positions of a single interval, as follows:
 for any $i,j\in I_t$, one has $i\prec j$ if either $i<j$ and $t$ is odd, or $i>j$ and $t$ is even.
This relation can be used to define the position \Def{ directly below} $i$, i.e. the closest position $i^-$ to $i$ such that $i^-\prec j$. In other words,  one has $i^-=i-1$ if $i,i-1$ belong to an odd interval, and $i-1\in I_t$ or $i^-=i+1$ if $i, i+1$ belongs to an even interval. In the absence of such a position, we set $i^-= 0$.
 The highest position of an interval $I_t$ is the position $i \in I_t$ such that there is no position $j$ in $I_t$ with $i\prec j$.
Now we can define the upward graph:
\begin{definition}
Given a wave embedding $W$ of an arc-annotated sequence $A=(S,P)$, the \textbf{upward graph} of $A$ associated to $W$ is the directed graph $G=(V_G, A_G)$ such that $V_G=P$, and $A_G$ is the set of arcs $((i,j) \mapsto (i',j'))$ such that $i$ or $j$ is directly below $i'$ or $j'$ in $W$\Alain{Formulation pas tre claire.}.\\
A wave embedding of an arc-annotated sequence $A=(S,P)$ is \textbf{ordering} if its upward graph is acyclic. 
\end{definition}

\begin{figure}
    \centering
     \includegraphics[width=.9\textwidth]{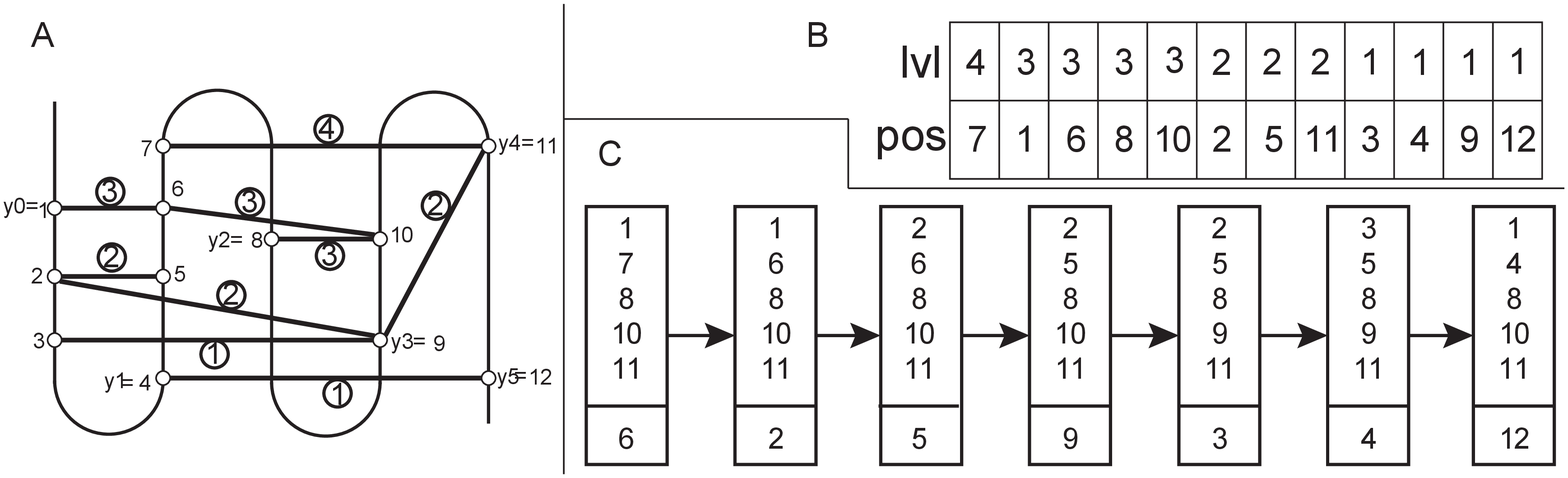}
 \caption{A: Arc-annotated sequence and its pivots. B: Ranking of the positions by decreasing for the level. C: Tree decomposition obtained with Algorithm~\ref{naive}. The highlighted position in each bag is the position denoted as position $p$ within Algorithm~\ref{naive}. The last position in each bag is the position $p^-$. }
 \label{treedecUpward}
\end{figure}

Algorithm~\ref{level} takes as input an upward graph, supposed to be acyclic, and assigns a level for each vertex, thus a level for each interaction of the associated arc-annotated sequence . This algorithm is a straightforward modification of Kahn's topological ordering algorithm~\cite{Kahn1962}, illustrated in~Figure \ref{treedecUpward}.

\begin{algorithm}[t]
  \SetKwInOut{Input}{Input}\SetKwInOut{Output}{Output}
  \Input{a directed acyclic graph $G=(V_G,A_G)$}
  \Output{Assign a level for each vertex}
\begin{itemize}
  \item $L=\{v \in V_G,~d^-(v)=0\}$\\
  \item \textbf{For each} $v \in L$, $level(v)=1$
  \item \textbf{While} $L \neq \varnothing$:
  \begin{itemize}
    \item pop front $v$ from $L$.
    \item \textbf{For each} $v'$ such that $(v,v') \in A_G$:
      \begin{itemize}
	\item remove $(v,v')$ from $A_G$
	\item \textbf{If} $d^-(v')=0$:
	  \begin{itemize}
	   \item push back $v'$ into $L$
           \item $level(v')=level(v)+1$
	  \end{itemize}
      \end{itemize}
  \end{itemize}
\end{itemize}
\caption{Level Algorithm \label{level}}
\end{algorithm}

Now we define the level of any position as the minimum level of all interactions in which the position is implicated, as  illustrated in Figure~\ref{treedecUpward}.
\begin{definition}[Level of a position]
 Given an ordering wave embedding $W$ of an arc-annotated sequence $A=(S,P)$, the \textbf{level of a position} $i$ defined by: $level(i) = \displaystyle \min_{(i,j) \in P}(level((i,j)))$. We define a \textbf{total order} $\curlyeqprec$ on $S$ through: $i \curlyeqprec j$ iff either $level(i) < level(j)$, or $level(i) = level(j)$ and $ i < j $.
\end{definition}

Then, we introduce Algorithm~\ref{naive} which, starting from an ordering wave embedding, decomposes any arc-annotated sequence. The key idea is to create a root which contains the highest position in each interval. The successor of a bag is then obtained by changing the highest level position into the position directly below it (Figure \ref{treedecUpward}).

\begin{algorithm}[t]
  \SetKwInOut{Input}{Input}\SetKwInOut{Output}{Output}
  \Input{an arc-annotated sequence $A$ and an ordering wave embedding of degree $k$}
  \Output{A tree decomposition of $A$}
\begin{itemize}
  \item Assign a level for each interaction using Algorithm \ref{level}, and map level to each position
  \item $X=\varnothing$ and $T$ is an empty tree
  \item Create a node $X_0$ composed of the highest position of each interval and set $X_0$ as the root of $T$
  \item $l=0$
 \item \textbf{While} there is a position $i\in X_l$ such that $i^-\neq 0$
    \begin{itemize}
      \item Search the position $p \in X_l$ with the highest level and such that $p^-\neq 0$
      \item Add $p^-$ to $X_l$
      \item Add $X_l$ to $X$
      \item If $l>0$, set $X_l$ as the son of $X_{l-1}$
      \item Set $l=l+1$ and $X_l= X_{l-1} - \{p\}$
    \end{itemize}
  \item \textbf{Return} $(X,T)$
\end{itemize}
\caption{Chaining Algorithm \label{naive}}
\end{algorithm}

\begin{theorem}
 Given an ordering wave embedding of degree $k$ for an arc-annotated sequence $A$, then a tree decomposition of $A$ having width $k$ can always be computed in time $O(k\cdot n)$.
\end{theorem}

\ShowProof{\begin{Proof}
We proved that given a wave embedding of degree $k$, we can build a smooth tree decomposition of width $k$.
The time complexity of the algorithm \ref{level} is obviously linear in the number of positions and interactions of the arc-annotated sequence.
The time complexity of the algorithm \ref{naive} is $O(kn)$ in terms of affectations, with $n$ the number of positions.
For each created bag, we make $k+1$ affectation, and we create $O(n)$ bag (we created exactly $n-k$ bags).
The searches of the highest position in the bags can be done at worst (and non efficiently) in $O(kn)$ as a search of the maximum in a non ordered list.
As for RNA, we have a constant maximum number of interactions for each position, the number of interaction $|P|=O(n)$.
So the final time complexity of the tree decomposition is $O(kn)$.
\qed \end{Proof}}

\begin{corollary}
  Let $A$ and $B$ be two arc-annotated sequences with $P_B= \varnothing$. Given an ordering wave embedding of degree $k$ of $A$, the structure-sequence alignment of $A$ and $B$ can be computed in $O(n\cdot m^k)$. 
\end{corollary}


\section{Application to three general classes of structures }

In this section, we define three new structure classes, which respectively generalize the standard pseudoknots \cite{Han2008}, the simple non-standard pseudoknots \cite{Wong2011} and the standard triple helices \cite{Wong2012}. For each of them, a \emph{natural} ordering wave embedding can be found, such that our general alignment algorithm has the same complexity as its, previously introduced, \emph{ad hoc} alternatives.

\subsection{Standard Structures}

Here we define and describe the alignment of \textbf{standard structures}, a natural generalization of the standard pseudoknots defined by Han {\it et al}~\cite{Han2008}. The main specificity of this class is that bases can interact with several other bases. This allows the consideration oft multiple non-canonical interactions (e.g. base triples) in RNA structures 
(see Figure \ref{all_struct}.A).
\begin{definition}[Standard Structure]
An arc-annotated sequence $A=(S,P)$ is a \textbf{standard structure} if there exists an ordering wave embedding, based on a pivot  list $\mathbf{y}=\{y_i\}_{i=0}^{k}$, $k>1$, such that the extremities of any interaction $(i,j) \in P$ are separated by exactly one pivot.

\end{definition}

The ordering wave embedding can then be used by Algorithm~\ref{naive} to yield a smooth tree decomposition of width $k$, therefore the complexity of the structure-sequence alignment is $O(n\cdot m^k)$.


\subsection{Simple Non-Standard Structures}

In \cite{Wong2011}, the algorithm of \cite{Han2008} is extended to capture the so-called simple non-standard pseudoknots.
Briefly, a simple non-standard pseudoknot contains a standard pseudoknot, and defines a special region from which interactions may initiate, possibly crossing interactions in the standard pseudoknot.
We extend this class in order to capture multiple interactions, as illustrated by Figure~\ref{all_struct}.B.
%
\begin{definition}[Simple Non-Standard Structure]
An arc-annotated sequence $A=(S,P)$ is a \Def{simple non-standard structure} (Type I) if there exist an ordering wave embedding, based on a pivot list $\mathbf{y}=\{y_i\}_{i=0}^{k}$, $k>1$ and $\tau\in [1,k-k'-1]$, $k'\in\{1,2\}$, such that the extremities of any interaction $(i,j) \in P$ with $j<y_{k-k'}$ are separated by exactly one pivot and the others interactions $(i,j) \in P$ with $y_{k-k'} \leq j$ are such that $y_{\tau-1} \leqslant i < y_{\tau}$.
\end{definition}
As in \cite{Wong2011}, Type II simple non-standard structures are symmetric to Type I:  the special region lies on the beginning of the sequence.
To be coherent with the definition of simple non-standard pseudoknots given in \cite{Wong2011}, we define the \Def{degree} of a standard structure as it number of pivots in its ordering wave embedding. Therefore, the treewidth of a simple non-standard structure of degree $k+1$ is at most $k+1$ and, given its pivots sequence, we can build a smooth tree decomposition of width $k+1$. Hence the complexity of the structure-sequence alignment is $O(n\cdot m^{k+1})$.

\subsection{Extended Standard Triple Helices}
To our knowledge, the standard triple helices~\cite{Wong2012} constitute the first attempt to handle base triples in sequence/structure alignments.
A standard triple helix is a kind of standard pseudoknot of degree $3$ where some positions are allowed to be involved in multiple base pairs.

\begin{figure}[t]
\centering
\begin{minipage}{0.7\textwidth}
  \includegraphics[width=\textwidth]{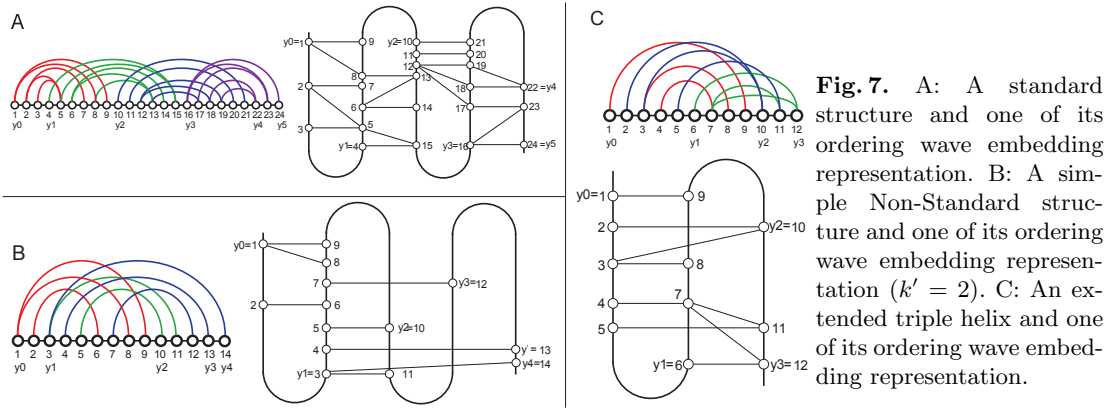}
 \label{alignmentEx}
\end{minipage}
\begin{minipage}{0.25\textwidth}
 \caption{A: A standard structure and one of its ordering wave embedding representation.
          B: A simple Non-Standard structure and one of its ordering wave embedding representation ($k'=2$).
	  C: An extended triple helix and one of its ordering wave embedding representation.}
 \label{all_struct}
\end{minipage}
\end{figure}

We define the \Def{extended standard triple helix} as the structures admitting an ordering wave embedding of degree $3$ (Figure \ref{all_struct}.C).  This new class strictly includes standard triple helices. 
Furthermore, each such structure  can be represented by a tree-decomposition which is smooth and has width at most $3$. This gives an algorithm in $O(n\cdot m^3)$ for the structure-sequence alignment.

\section{Recursive Structures}

Now we consider much more general RNA structures, where different kinds of pseudoknots can occur anywhere. As will be seen below, such structures can be decomposed into \Def{primitives}, and from the tree-decomposition of each primitive a global tree-decomposition of the structure can be built. The set of \Def{primitive sub-arc-annotated subsequences} (\Def{primitives} for short) of an arc-annotated sequence is the set of all sub-arc-annotated sequences induced by the connected components of its \Def{conflict graph}, which is defined as follows.
 The conflict graph $G=(V,E)$ of an arc-annotated sequence $A=(S,P)$ is the graph such that:\setlist{nolistsep}
  \begin{itemize}
  \item $V=P$ (the nodes of $G$ are the interactions of $A$).
  \item $(v_1,v_2) \in E$ with $v_1=(i_1,j_1)$ and $v_2=(i_2,j_2)$ ($i_1<i_2$) iff $i_1 < i_2 < j_1 < j_2$ (interactions cross).
  \end{itemize}
The \textbf{boundaries} of a primitive are its left-most and right-most positions.

Let $A$ and $A'$ be two primitives of an arc-annotated sequence, and let $i'$ and $j'$ be the boundaries of $A'$ ($i'<j'$). We say that $A$ is \Def{encapsulated} in $A'$ iff for any position $i\in A$, one has $i' \leq i \leq j'$, and there exists at least one position $j\in A$ such that $ i' < j < j'$.
The \Def{depth} of a primitive of an arc-annotated sequence is the number of primitives which encapsulate it. We say that $A$ is \Def{directly} encapsulated in $A'$ if $A$ is encapsulated in $A'$ and $depth(A)=depth(A')+1$.

The \textbf{extension} of a primitive $A$ of depth $i$ is the arc-annotated subsequence consisting of: the primitive $A$; the   
boundaries of any primitive that is directly encapsulated in $A$; and the unpaired positions that are directly encapsulated in $A$.

%
Given a primitive $A_0$ of depth $0$ of an arc-annotated sequence $A$, the \Def{next primitive} of $A_0$ is the following primitive of level $0$ in the sequential order (note that they can share a boundary).

\begin{algorithm}[t]
\begin{itemize}
 \item Compute a tree decomposition for all primitive extensions using Algorithm \ref{naive}.
 \item For each primitive $A_0$ of depth $0$:
    \begin{itemize}
    \item Create a bag $\chi$ containing the right boundary of $A_0$ and the left boundary of its next primitive $A'_0$.
    \item Let $(X^0, T^0)$ be the tree decomposition of the extension of $A_0$ and $(X'^0, T'^0)$ the one of the extension of $A'_0$.
    \item Add to $\chi$ the position $i$ of the root of $(X'^0, T'^0)$ such that $i-1$ or $i+1$ belongs to the root too.
    \item Connect the leaf of $(X^0, T^0)$ to $\chi$ and connect $\chi$ to the root of $(X'^0, T'^0)$.
    \end{itemize}
 \item For each possible depth $i>0$ in increasing order:
 \begin{itemize}
  \item For each primitive $A$ of depth $i$:
    \begin{itemize}
    \item Let $(X,T)$ be the tree decomposition of the extension of $A$ and $(X', T')$ the one of the extension of the arc-annotated sequence $A'$ in which $A$ is encapsulated.
    \item Find a bag $\chi$ in $(X', T')$ such that $\chi$ contains the boundaries of $A$.
    \item Connect $(X,T)$ to $(X',T')$ by connecting the root of $(X, T)$ to $\chi$.
    \item Add the right-most boundary of $A$ in $(X,T)$ to all bags from the root to the first bag containing it.
    \end{itemize}
  \end{itemize}
\end{itemize}
\caption{Recursive Algorithm \label{recursive}}
\end{algorithm}


\begin{figure}[t]
 \centering
 \includegraphics[width=0.9\textwidth]{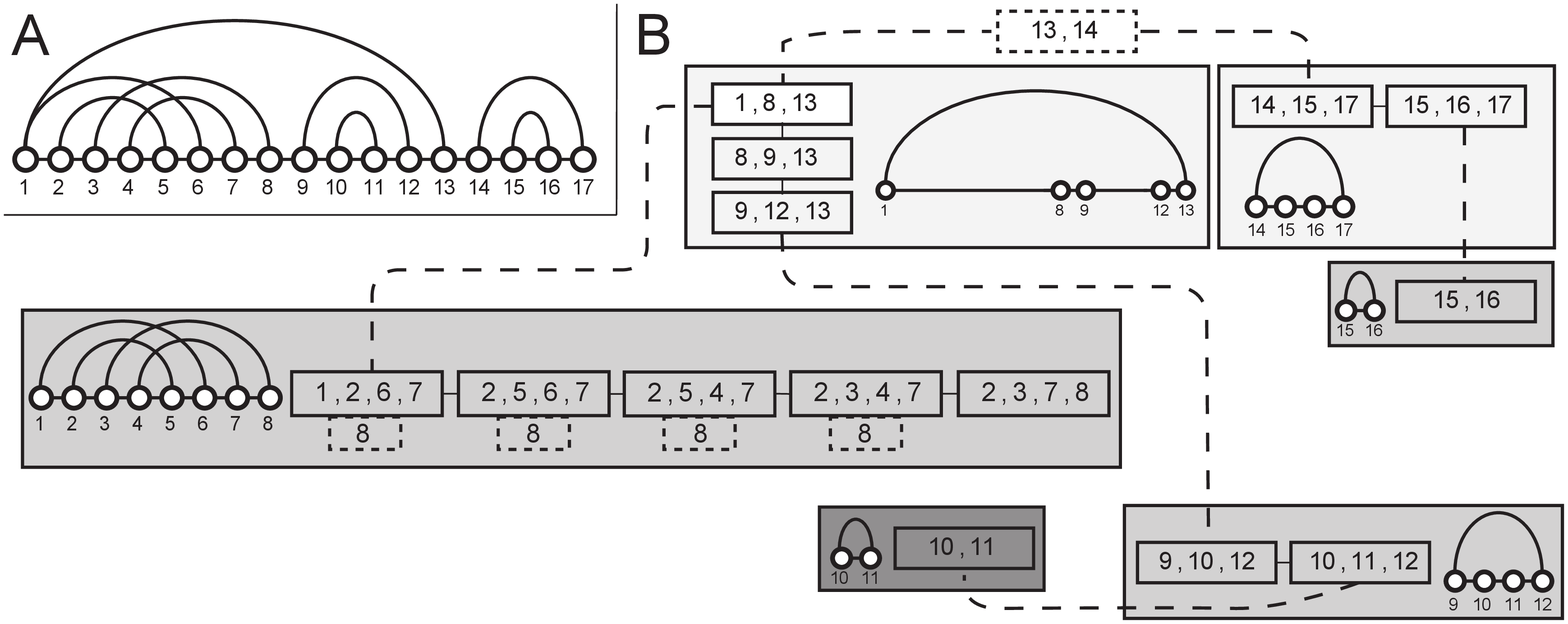}
 \caption{(A) An arc-annotated sequence and (B) a representation of the tree decomposition given by Algorithm \ref{recursive}.
	  Each box correspond to an extension of a primitive and its associated tree decomposition.
          Dashed links correspond to the connections made by Algorithm \ref{recursive}.
          The Dashed bag (top of B) correspond to the bag $\chi$ added to connect two consecutive primitive of level $0$.
          The position $8$ in a dashed box illustrates the case where the right boundary of a primitive need to be added in its tree decomposition.}
 \label{rec_tree_dec}
\end{figure}

\begin{theorem}
Let $A$ be an arc-annotated sequence of size $n$. If there exist an ordering wave embeddings of degree at most  $k$ for each extension 
of its primitives, then the treewidth of $A$ is at most $k+1$, and a $\kappa$-weakly smooth tree decomposition of $A$ can be built in $O(k\cdot n)$ time, where $\kappa$ is the number of primitives of odd degree, whose level is greater or equal to $1$.\label{th:rec}
\end{theorem}

\begin{corollary}
Let $A$ and $B$ be two arc-annotated sequences with $P_B= \varnothing$. Given an ordering wave embedding of degree $k$ (at most) for each extension of the primitives of $A$, the structure-sequence alignment of $A$ and $B$ can be computed in $O(\kappa \cdot m^{k+1} + n \cdot m^k)$ (with $\kappa$ defined in Theorem~\ref{th:rec}).
\end{corollary}

\section{Conclusion}

We have given a general parameterized dynamic programming scheme for sequence-structure comparison in a large class of RNA structures, which unifies and generalizes several families of structures that have been independently considered by previous works. Notably, we can handle structures where each nucleotide can be paired to any number of other nucleotides, thus cqpturing any type of non-canonical interactions. Our approach relies on a tree decomposition approach of arc-annotated sequences represented as wave embeddings, and the treewidth of the decomposition is then equal to the degree of the wave embedding. Computing a wave embedding of small degree is easy for all classes of pseudoknotted structures considered in this paper. However, the problem of finding a minimum degree wave embedding for any kind of pseudoknotted structure remains open.

\section{Acknowledgments}
This work was supported by the {\sc DIGITEO} RNAOmics project (AD, PR and YP), the {\sc ANR} AMIS ARN project (ANR-09-BLAN-0160, AD and PR) and the {\sc ANR} MAGNUM project (ANR-2010-BLAN-0204, YP).

\bibliographystyle{splncs03}
\bibliography{ss_alignment}


\end{document}